\documentclass{INTERSPEECH2023}

\usepackage{amsmath,amssymb,graphicx}
\usepackage{mathrsfs}
\usepackage{multirow}
\usepackage{booktabs}
\usepackage{stfloats}
\usepackage{cite}
\usepackage{enumitem}
\usepackage{threeparttable}

\interspeechcameraready

\title{Language-Routing Mixture of Experts for Multilingual and Code-Switching Speech Recognition}

\name{Wenxuan Wang, Guodong Ma, Yuke Li$^*$\thanks{$^*$ Corresponding author}, Binbin Du}

\address{NetEase Yidun AI Lab, Hangzhou, China}

\email{{$\left\{wangwenxuan,maguodong,liyuke,dubinbin\right\}$@corp.netease.com}}

\begin{document}

\maketitle
 
\begin{abstract}
Multilingual speech recognition for both monolingual and code-switching speech is a challenging task. Recently, based on the Mixture of Experts (MoE), many works have made good progress in multilingual and code-switching ASR, but present huge computational complexity with the increase of supported languages. In this work, we propose a computation-efficient network named \textbf{L}anguage-\textbf{R}outing \textbf{M}ixture \textbf{o}f \textbf{E}xperts (LR-MoE) for multilingual and code-switching ASR. LR-MoE extracts language-specific representations through the Mixture of Language Experts (MLE), which is guided to learn by a frame-wise language routing mechanism. The weight-shared frame-level language identification (LID) network is jointly trained as the shared pre-router of each MoE layer. Experiments show that the proposed method significantly improves multilingual and code-switching speech recognition performances over baseline with comparable computational efficiency.
\end{abstract}

\vspace{-0.3 em}
\section{Introduction}
\label{sec:intro}
\vspace{-0.1 em}

Multilingualism is a widespread phenomenon in the world. Multilingual speakers often communicate in multiple languages simultaneously, such as interspersing English in Mandarin. Therefore, a practical multilingual speech recognition system needs to support the recognition of monolingual and code-switching utterances in multiple languages.

End-to-end (E2E) ASR systems\cite{graves2006connectionist,graves2012sequence,graves2013speech,2015Attention,kim2017joint,2016Listen,hori2017advances,ma21_interspeech,ma22_interspeech} have become more and more popular recently due to the simple pipeline, excellent performance and less dependence on linguistic knowledge compared to traditional hybrid methods\cite{2012Deep}. Prior works based on the E2E model have also made good progress in the field of multilingual ASR, including code-switching corpus synthesis\cite{2018Language,2019Linguistically,2019Comparison}, multi-task training with joint language identification\cite{luo2018towards,kannan2019large}, self-supervised speech representation learning\cite{conneau2020unsupervised,jacobs2021acoustic,lahiri2021multilingual}, cross-lingual transfer learning\cite{kannan2019large,2021Meta}, etc. 
The MoE architecture is an effective method to improve the performance of multilingual speech recognition in both monolingual and code-switching scenarios, which has been widely concerned and studied recently. The existing MoE-based methods \cite{dalmia2021transformer,2020Bi,yan2022joint,Song2022LanguagespecificCA,tian22c_interspeech,9414379,mole_paper} extract language-specific representations separately by independent encoders and fuse them to decode. Mostly, the computational complexity of the models will increase significantly with the number of supported languages. 

In this work, we propose a computation-efficient network named \textbf{L}anguage-\textbf{R}outing \textbf{M}ixture \textbf{o}f \textbf{E}xperts (LR-MoE) to improve the performance of the multilingual and code-switching ASR task. The LR-MoE architecture consists of a shared block and a Mixture-of-Language-Experts (MLE) block. 
Unlike the sparsely-gated mixture of experts (sMoE) \cite{shazeeroutrageously,fedus2022switch,you2021speechmoe, Kumatani2021BuildingAG}, the expert layers in the MLE block are language-dependent, which is called Language-Specific Experts (LSE). 
The shared block generates the global representation, while the LSE of the MLE block extracts language-specific representations. In the MLE block, we design a Frame-wise Language Routing (FLR) mechanism, which guides the expert layers to learn language specialization at the training stage. A weight-shared frame-level language identification (LID) network is jointly trained as the shared pre-router of each LSE layer, and the alignment of frame-wise LID will be used as the routing path of the LSE layers. We also compare utterance-wise and frame-wise language routing for LR-MoE in the multilingual and code-switching experiment. To distinguish them, we will name the two networks with different routing as ULR-MoE and FLR-MoE, respectively.
Our contributions are summarized as follows:

\begin{itemize}[itemsep=2pt,topsep=3pt,parsep=3pt,leftmargin=15pt]
    \vspace{-0.25 em}
    \item We propose a computation-efficient LR-MoE architecture, which is suitable to apply in more languages with little increase in computational complexity.
    \vspace{-0.25 em}
    \item We investigate multiple routing strategies of MoE and propose an FLR mechanism to guide the expert layers to learn language specialization, which is compatible with both multiple monolingual and code-switched ASR.  
    \vspace{-0.25 em}
    \item In Mandarin-English code-switching and multilingual experiments, the proposed method significantly improves the performances of multilingual and code-switching speech recognition over the baseline with comparable computational efficiency and outperforms previous MoE-based methods with less computational complexity.
\end{itemize}


\vspace{-0.5 em}
\section{Related Works and Motivation}
\label{sec:pre}


\vspace{-0.3 em}
\subsection{Previous MoE-based works}
\label{ssec:bi-encoder}
\vspace{-0.3 em}

More recently, many works \cite{dalmia2021transformer,2020Bi,yan2022joint,Song2022LanguagespecificCA,tian22c_interspeech} focus on exploring MoE architectures to recognize monolingual and intra-sentence code-switching speech. The MoE-based methods mainly utilized language-specific expert encoders to generate parallel language-specific representations and fuse them, whose difference is primarily in the fusion mode of expert encoders and training strategy. For example, the Bi-encoder transformer network \cite{2020Bi} uses a gated network to dynamically output the MoE interpolation coefficients to mix two encoding representations. The weights of expert encoders are initialized with the pretrained monolingual model, respectively. Conditional factorized neural transducer \cite{yan2022joint} defined the monolingual sub-tasks with label-to-frame synchronization to achieve unified modeling of code-switching and monolingual ASR. The language-aware encoder \cite{tian22c_interspeech, Song2022LanguagespecificCA} learned language-specific representations through language-aware training with the language-specific auxiliary loss instead of monolingual pretraining and used the frame-wise addition to fuse them. 

\subsection{Motivations}
\label{ssec:rethink}

As mentioned above, the previous MoE-based works achieved considerable improvement in monolingual and code-switching ASR, but there are still the following problems:
\begin{itemize}[itemsep=2pt,topsep=3pt,parsep=3pt,leftmargin=15pt]
    \item The approach needs to compute all language-specific blocks. However, only one works in the monolingual scene. It means a large amount of redundant computational overhead. And the more languages are supported, the more redundant computational overhead is.
    \vspace{-0.3 em}
    \item Language-specific blocks are isolated from each other and lack interaction. As a result, the cross-linguistic contextual information is easily lost in code-switching speech.
\end{itemize}

In order to alleviate the above two issues, we propose the LR-MoE architecture inspired by the sparsely-gated mixture of experts \cite{shazeeroutrageously, fedus2022switch, you2021speechmoe, Kumatani2021BuildingAG}. Please refer to Section~3 for more details.



\vspace{-0.8 em}
\begin{figure}[htb]

\begin{minipage}[htb]{0.45\linewidth}
  \centering
  \centerline{\includegraphics[width=4.cm]{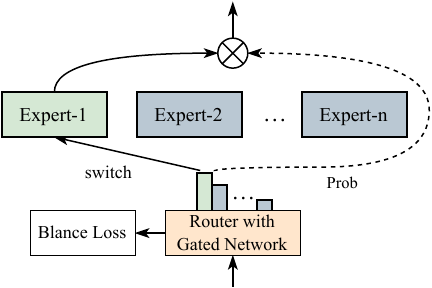}}
  \centerline{(a)} \medskip
\end{minipage}
%
\hfill
\begin{minipage}[htb]{0.45\linewidth}
  \centering
  \centerline{\includegraphics[width=4.cm]{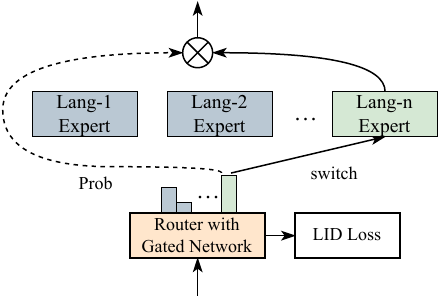}}
  \centerline{(b)} \medskip
\end{minipage}
\vspace{-1.0 em}
\caption{Schematic diagram of MoE modules. (a) Sparsely-Gated Mixture of Experts (sMoE), (b) Mixture of Language Experts (MLE).}
\vspace{-1.5 em}
\label{fig:module}
\end{figure}

\begin{figure}[t]
\centering
 \centerline{\includegraphics[width=8.0cm]{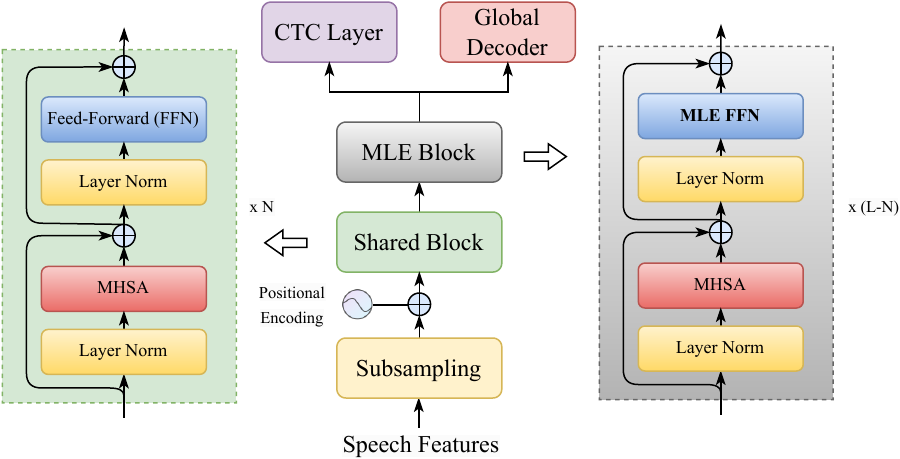}}
\vspace{-0.5 em}
\caption{The structure of the LR-MoE Transformer Model. $N$ and $(L-N)$ are the number of layers of the shared block and the MLE block, respectively.}
\vspace{-1.5 em}
\label{fig:arch}
\end{figure}

\vspace{-0.8 em}
\section{Proposed Method}
\label{sec:proposed}



\subsection{Sparsely-Gated Mixture of Experts}
\label{ssec:sparse-moe}

The sMoE module is shown in Fig.~\ref{fig:module}(a). As a representative, Switch Transformer \cite{fedus2022switch} adopts a top-1 expert routing strategy in the MoE architecture to route the data samples to the expert model with the highest probability in the gated network. The computational complexity of the whole network increases slightly as the number of experts increases, and the extra computational overhead only comes from the gated network. The inputs of the expert layer and the gated network are the outputs of the previous non-expert layer $o_{ne}$. The router probability $p$ can be expressed as follows:
\begin{equation}\label{eq1}
p = Softmax(W_{r} \cdot o_{ne} + b_{r}) \tag{1}
\end{equation}
where $W_{r}$ and $b_{r}$ is the weights and bias of router respectively.
An auxiliary loss is added to guarantee load balance across the experts during training. The balance loss is expressed as:
\begin{equation}\label{eq2}
\mathcal{L}_{b} = n \cdot \sum_{i=1}^{n}f_{i} \cdot p_{i} \tag{2}
\end{equation}
where $f_{i}$ is the fraction of samples dispatched to i-th expert, $n$ is the number of experts.

\vspace{-0.5 em}
\subsection{Architecture of LR-MoE}
\label{ssec:lrmoe}
\vspace{-0.3 em}

In order to strengthen the interaction of cross-lingual context information, we further expand the shared parts, such as the attention layers in the Transformer \cite{vaswani2017attention} network. It is different from the separate language-specific encoder \cite{dalmia2021transformer,2020Bi,yan2022joint,Song2022LanguagespecificCA}. Fig. \ref{fig:arch} shows the LR-MoE-based Transformer model. The shared block are stacked by standard transformer blocks. In contrast to the standard transformer block, we introduce the MLE
FFN module as shown in Fig. \ref{fig:module}(b) to strengthen language-specific representations in MLE block. All MLE modules share the same language router, which is a frame-level LID-gated network in front. According to the top-1 language predicted by the FLR, data samples are routed to the corresponding LSE layer. For each time frame, only one LSE will be routed, so the computational complexity of the model will not increase with more languages.

\subsection{Language Routing}
\label{ssec:FLR}
\subsubsection{LID-Gated Network}
\label{sssec:lid-gate}

Considering that LID can be regarded as a low-dimensional subtask of ASR and the output of the non-expert layer $o_{ne}$ already contains rich high-dimensional linguistic information, we model the frame-level LID task with a linear layer as follows:
\begin{equation}\label{eq3}
r = W_{r} \cdot o_{ne} + b_{r} \tag{3}
\end{equation}


A frame-level LID auxiliary loss is added to jointly train the LID and ASR tasks at the training stage. We get the auxiliary labels ${\cal Y}_{\rm lid}$ by replacing the tokens in the text labels with the corresponding language IDs. Then, based on $r$, the LID-CTC loss is adopted to get the token-to-frame alignment, as shown in Eq.~(\ref{ctc_lid}). 
\vspace{-0.3 em}
\begin{equation}
    \label{ctc_lid}
    \mathcal{L}_{lid-ctc} = - \log P_{\rm CTC}({{\cal Y}_{\rm lid}} | r) \tag{4}
\end{equation}
\vspace{-0.5 em}


Due to the sparse spike property of the Connectionist Temporal Classification (CTC), the greedy decoding result of the LID output $z_{t}$ will contain a large amount of blank. Therefore, we adopt a simplified alignment strategy to get dense frame-wise language routing information as follows:
\begin{equation}\label{eq5}
    z_{t} =
    \begin{cases}
      z_{f}, & \mbox{if}~t = 0 \\
      z_{t}, & \mbox{if}~z_{t} \neq \phi \\
      z_{t-1}, & \mbox{if}~z_{t} = \phi
    \end{cases} \tag{5}
\end{equation}
where $z_{t} \in \{language\ ids\} \cup \phi,  t=0,1,2,…,T$, $z_{f}$ is the first non-blank element.

Besides, we also use an utterance-wise LID-gated network with the cross entropy (CE) loss for comparison. The utterance-wise LID loss is as follows:
\begin{equation}
    \label{utt_lid}
    \mathcal{L}_{lid-utt} = CE({r}_{u},\ {\cal U}_{\rm lid}) \tag{6}
\end{equation}
where ${r}_{u}$ is the time-dimension global average pooling of $r$, ${{\cal U}_{\rm lid}}$ is the language ID of utterance.

The final object loss function is shown in Eq.~(\ref{eq7}):
\begin{equation}\label{eq7}
\mathcal{L}_{mtl} = \mathcal{L}_{asr} + \lambda_{lid} \mathcal{L}_{lid} \tag{7}
\end{equation}
where $\lambda_{lid}$ is selected by hand, $\mathcal{L}_{lid}$ of ULR and FLR correspond to $\mathcal{L}_{lid-utt}$ and $\mathcal{L}_{lid-ctc}$, respectively.

\subsubsection{Shared Router}
\label{sssec:sr}

Unlike sMoE \cite{shazeeroutrageously,fedus2022switch,you2021speechmoe, Kumatani2021BuildingAG}, we use a shared router instead of the independent router for each MLE layer, mainly due to the following considerations:
The independent router of each MoE layer in sMoE is helpful in obtaining more diverse routing paths and larger model capability. However, the expert layers are language-specific and the desired routing paths are determined with a priori in LR-MoE. Therefore, the shared LID router might be helpful to reduce additional computation and the multi-level error accumulation caused by the alignment drift of the language routing.

\subsection{Pretrained Shared Block}
\label{ssec:pretrain}

The bottleneck features of ASR are effective in transfer learning for LID \cite{wang2021end}. Therefore, we utilize a pretrained shared block to speed up the convergence of the LID-gated network and reduce the bad gradient back-propagation caused by erroneous routing paths, especially at the early training stage.

\section{Experiments}
\label{sec:experiments}

\subsection{Datasets}
\label{ssec:setup}
Our experiments are conducted on ASRU 2019 Mandarin-English code-switching Challenge dataset\cite{shi2020asru} and a four-language dataset including Aishell-1 (CN) \cite{bu2017aishell}, train-clean-100 subset of Librispeech \cite{panayotov2015librispeech} (EN), Japanese (JA), Zeroth-Korean (KR) \footnote{https://openslr.org/40/} and Mandarin-English code-switching (CN-EN) data. JA and CN-EN are collected from Datatang \footnote{https://www.datatang.com/}. Table \ref{tab:asru data} and \ref{tab:multilingual data} show the details of all experimental datasets.

\vspace{-0.5 em}
\begin{table}[ht!] 
\setlength\tabcolsep{4.5pt}
  \caption{Details of Mandarin-English Code-Switching Dataset}
  \vspace{-1.5 em}
  \label{tab:asru data}
    \center
    \begin{tabular}{c c | c c | c c} \hline
    \multicolumn{1}{c}{\multirow{2}{*}{Lang}} & \multicolumn{1}{c|}{\multirow{2}{*}{Corpora}} & \multicolumn{2}{c|}{Dur. (Hrs)} & \multicolumn{2}{c}{Utterance(k)} \\ \cline{3-6}
      &   & Train & Eval & Train & Eval \\ 
    \hline
    CN & ASRU-Man\cite{shi2020asru} & 482.6 & 14.3 & 545.2 & 16.6\\ 
    EN & Librispeech\cite{panayotov2015librispeech} & 464.2 & 10.5 & 132.5 & 5.6 \\
    CN-EN & ASRU-CS\cite{shi2020asru} & 199.0 & 20.3 & 186.4 & 16.2\\
    \hline
    \end{tabular}
\end{table}


\vspace{-1.8 em}
\begin{table}[ht!] 
\setlength\tabcolsep{4.5pt}
  \caption{Details of Multilingual and Code-Switching Dataset}
  \vspace{-1.5 em}
  \renewcommand\tabcolsep{4.5pt}
  \label{tab:multilingual data}
    \center
    \begin{tabular}{ c | l |c c c c c} \hline
     Split & Information & CN& EN& JA & KR& CN-EN \\
    \hline
    \multicolumn{1}{c|}{\multirow{2}{*}{Train}} & Dur. (Hrs.) & 151.2& 100.6 & 93.7 & 51.6 & 93.1 \\
      & Utterances(k) & 120.1 & 28.5 & 75.5 & 22.3 & 84.0\\
    \hline
    \multicolumn{1}{c|}{\multirow{2}{*}{Eval}} & Dur. (Hrs.) & 9.7&  5.4 & 6.3 & 1.2 & 6.9 \\
      & Utterances(k) & 7.2 & 2.6 & 5.2 & 0.5 & 6.6 \\
    \hline
    \end{tabular}
\end{table}


\vspace{-0.6 em}

For all the experiments, the acoustic features are 80-dimensional log filter-bank energy extracted with a stride size of 10ms and a window size of 25ms. SpecAugment\cite{park2019specaugment} is applied during all training stages. The Mandarin-English vocabulary and the vocabulary of 4 languages consist of 12064 and 15492 unique characters and BPE \cite{sennrich2015neural} tokens.

\subsection{Experimental setup}
\label{ssec: setup}

The experiments are conducted on the ESPnet toolkit \cite{watanabe2018espnet}. We use the Transformer CTC/Attention model with a 12-layer encoder and a 6-layer decoder as our baseline, called the Vallina model. The LR-MoE encoder consists of a 6-layer shared block and a 6-layer MLE block in experiments. 
We also compare various MoE-based encoders with our proposed method, including Bi-Encoder \cite{2020Bi}, LAE\cite{tian22c_interspeech}. The Bi-Encoder contains the 12-layer encoder for each language. The LAE contains a 9-layer shared block and a 3-layer language-specific block for each language. Besides, we implement 12-layer sMoE\cite{Kumatani2021BuildingAG} with 4 experts in each MoE block for comparison in the multilingual and code-switching experiment. 
All encoders and decoders are stacked transformer-based blocks with attention dimension of 256, 4 attention heads and feed-forward dimension of 2048.
We implement multi-task learning with $\lambda_{ctc}=0.3$ and $\lambda_{lid}=0.3$ for ASR and LID at the training stage of the LR-MoE model.

We use the Adam optimizer with a transformer-lr scale of 1 and warmup steps of 25k to train 100 epochs on 8 Tesla V100 GPUs. The training process adopts a dynamic batch size strategy with a maximum batch size of 128. We train a 4-gram language model with all training transcriptions and adopt the CTC prefix beam search for decoder with a fixed beam size of 10.

\subsection{Experimental Results}
\label{ssec: result}

\subsubsection{Results on Mandarin-English ASR}
\label{ssec: Mandarin-English result}

\begin{table}[ht!] 
  \renewcommand\tabcolsep{4.5pt}
  \begin{threeparttable}
  \caption{\textbf{Performance of models in the CTC-based and AED-based Mandarin-English ASR system\tnote{*}.} "CN", "EN" and "ALL" mean the character error rate (CER) of monolingual Mandarin, the word error rate (WER) of monolingual English and the total mix error rate (MER) of code-switching test set respectively.} 
  \vspace{-1.5 em}
  
  \label{tab:asru}
    \center

    \begin{tabular}{l c | c c | c c c} \hline
    \multicolumn{1}{c}{\multirow{2}{*}{Model}} &
    \multicolumn{1}{c|}{\multirow{2}{*}{Params}} &
    \multicolumn{2}{c|}{Mono} & \multicolumn{3}{c}{Code-Switch}\\ \cline{3-7}
     & & CN & EN & ALL & CN & EN \\ \hline  
    Vallina CTC & 19.8M & 7.1 & 12.4 & 12.2 & 9.0 & 38.9 \\
    Bi-Encoder CTC & 36.6M & 5.3 & 10.3 & 10.7 & 7.9 & 33.3 \\
    LAE CTC & 26.5M & 5.3 & 10.5 & 10.8 & 8.0 & 33.7 \\
    FLR-MoE CTC & 25.8M &  \textbf{5.1} & \textbf{10.1} & \textbf{10.5} & \textbf{7.7} & \textbf{33.1} \\
    \hline
    Vallina AED & 34.7M &  6.3 & 11.7 & 11.2 & 8.6 & 32.5 \\
    Bi-Encoder AED & 51.5M & 4.9 & 9.8 & 9.9 & 7.6 & 28.9 \\
    LAE AED & 41.4M &  5.0 & 9.9 & 10.0 & 7.7 & 29.2  \\
    FLR-MoE AED & 40.7M &  \textbf{4.7} & \textbf{9.6} & \textbf{9.7} & \textbf{7.4} & \textbf{28.4} \\
    \hline    
    \end{tabular}
    
    \begin{tablenotes}    
        \footnotesize               
        \item[*] Note: The results of the Bi-Encoder and LAE are achieved with our experimental configuration.        
      \end{tablenotes}            
    \end{threeparttable}
    
\end{table}
\vspace{-0.5 em}

As shown in Table \ref{tab:asru}, our proposed method outperforms the previous MoE-based methods with comparable or fewer parameters in the Mandarin-English ASR system, including CTC and Attention-based Encoder-Decoder (AED) models. The proposed method achieves significant performance improvement over the baseline. The relative improvements on mono-Mandarin, mono-English and code-switch evaluation sets are 28.2\%, 18.5\% and 13.9\% in the CTC-based model and 25.4\%, 17.9\% and 13.4\% in the AED-based model, respectively.

\begin{table*}[ht!] 
  \renewcommand\tabcolsep{4.5pt}
  \caption{\textbf{Performance of models in the CTC-based multilingual ASR system.} Multi-Encoder is a multilingual extension of the method of Bi-Encoder\cite{2020Bi} with the
unsupervised gating network. 
"CN", "EN", "JA", "KR" and "ALL" mean CER of Mandarin, WER of English, CER of Japanese, CER of Korean and MER of Mandarin-English code-switch, respectively. We also use the number of parameters and floating-point operations (FLOPs) of a 30s input to evaluate the computational complexity of the model.}
  \vspace{-1.5 em}
  
  \label{tab:cn-en-ja-kr}
    \center
    \begin{tabular}{l c c c| c c c c c | c c c } \hline
    \multicolumn{1}{c}{\multirow{2}{*}{Model}} &
    Shared &
    \multicolumn{1}{c}{\multirow{2}{*}{Params}} &
    \multicolumn{1}{c|}{\multirow{2}{*}{GFLOPs}} &
    \multicolumn{5}{c|}{Monolingual} & \multicolumn{3}{c}{Code-Switch}\\ \cline{5-12}
      & Router & & & CN(CER) & EN(WER) & JA(CER) & KR(CER) & Avg & ALL(MER) & CN & EN \\ \hline
    Vallina CTC & - & 19.8M & 55.4 & 6.3 & 13.1 & 10.1 & 5.6 & 8.8 & 10.4 & 6.9 & 29.7 \\
    Multi-Encoder & - & 71.0M & 146.9 & 5.9 & 10.9 & 9.1 & 2.6 & 7.1 & 9.4 & 6.6 & 24.2 \\
    LAE & - & 35.7M & 78.3 & 5.6 & 11.3 & 8.8 & 2.4 & 7.0 &8.6 & 6.1 & 22.5 \\
    sMoE & w/o & 57.4M & 55.4 & 5.5 & 10.7 & 8.6 & 3.2 & 7.0 &9.3 & 6.5 & 25.1 \\
    ULR-MoE & w/o &  38.6M & 55.4 & 5.3 & 10.4 & 8.4 & 2.0 & 6.5 & 8.6 & 5.9 & 23.5 \\
    FLR-MoE  & w/o & 38.6M & 55.4 & 5.2 & 10.5 & 8.5 & 2.1 & 6.6 & 8.4 & 5.9 & 22.4 \\
    FLR-MoE (ours)  & w & 38.6M & 55.4 & \textbf{5.2} & \textbf{10.0} & \textbf{8.2} & \textbf{1.9} & \textbf{6.3} & \textbf{8.2} & \textbf{5.8} & \textbf{21.8} \\
    \hline
    \vspace{-2.0 em}
    \end{tabular}
\end{table*}

\subsubsection{Results on multilingual ASR}
\label{ssec: multilingual result}



Table \ref{tab:cn-en-ja-kr} shows the results of models in the CTC-based multilingual ASR system. Compared with previous MoE-based methods, our proposed method achieves significant performance improvement in both monolingual and code-switching scenarios. Regarding FLOPs, the proposed architecture's computational complexity increases little with the increase of languages, and it is easier to scale in multilingual ASR. 

The proposed method significantly improves performance over the baseline with comparable computational complexity. The relative average improvements on monolingual and code-switch evaluation sets are 28.4\% and 26.8\%, respectively. 

We also compare different routing strategies of LR-MoE. Experiments show that FLR-MoE achieves slight performance improvement compared to ULR-MoE, especially in code-switching scenarios. The shared language routing strategy for FLR-MoE outperforms the layer-wise language routing strategy on both monolingual and code-switching evaluation sets.

\subsection{Ablation study and analysis}
\label{ssec: ab}

\subsubsection{Position of MLE}
\label{sssec: lse position}

\begin{table}[ht!] \small
  \setlength{\abovecaptionskip}{-0cm} 
  \setlength{\belowcaptionskip}{-0.2cm}
  \caption{\textbf{Ablation study on the position of the MLE FFN layers.} We compare the average token error rate (TER) of four monolingual evaluation sets, the total mix error rate (MER) of the code-switching evaluation set and the average utterance-level LID accuracy for the different positions of the MLE module.}
  \vspace{-0.5 em}
  \label{tab:position}
    \center
    \begin{tabular}{c c c c c} \hline
    Position & Shared & Monolingual & Code-Switch & LID \\ 
    of MLE & layers & (Avg TER) & (MER) & Acc \\
    \hline
     - & 12 & 8.8 & 10.4 & 98.8   \\
    \{10-12\} & 9 & 7.0 & 8.4 & \textbf{99.5}   \\
    \{7-12\} & 6 &\textbf{6.3} & \textbf{8.2} & 99.4 \\ 
    \{4-12\} & 3 & 6.8 & 9.3 & 99.1 \\ 
    \{1-12\} & 0 & 11.6 & 16.2 & 97.7 \\ \hline
    \end{tabular}
\end{table}

\vspace{-0.5 em}

As shown in Table \ref{tab:position}, we explore the effect of the location the position of the MLE layers on performance. According to our analysis, the deeper the shared block, the more accurate the LID and the weaker the language-specific representation in the fix-depth model of FLR-MoE, which is a trade-off of the model design. Experiments show the proposed method has a strong ability to distinguish languages, and we achieve the best results with the language router in the middle position.

\subsubsection{LID and Routing Analysis}
\label{sssec: lid analysis}

\begin{figure}[t]
\centering
 \centerline{\includegraphics[width=8.0cm]{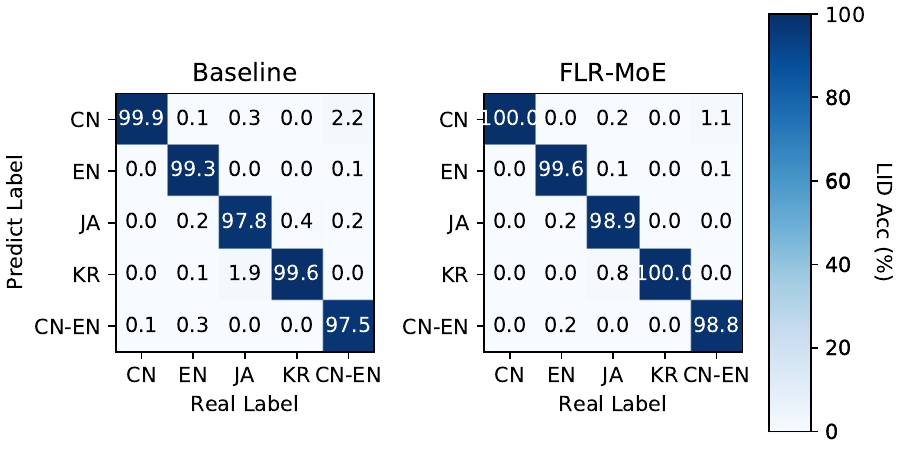}}
\vspace{-0.5 em}
\caption{Confusion matrix of language over the multilingual evaluation sets. left: Baseline, right: FLR-MoE.}
\vspace{-0.5 em}
\label{fig:confuse}
\end{figure}

We summarize the results of utterance-level language classification based on the frame-level LID routing information and the results of ASR. The proposed method's language classification accuracies and confusion matrix are shown in Fig. \ref{fig:confuse}. Compared to the 98.8\% average LID accuracy of the baseline, the proposed method's average LID accuracy is 99.4\%. This shows that the proposed method can effectively reduce the confusion between languages.

\begin{figure}[htb]
\centering
 \centerline{\includegraphics[width=7.5cm]{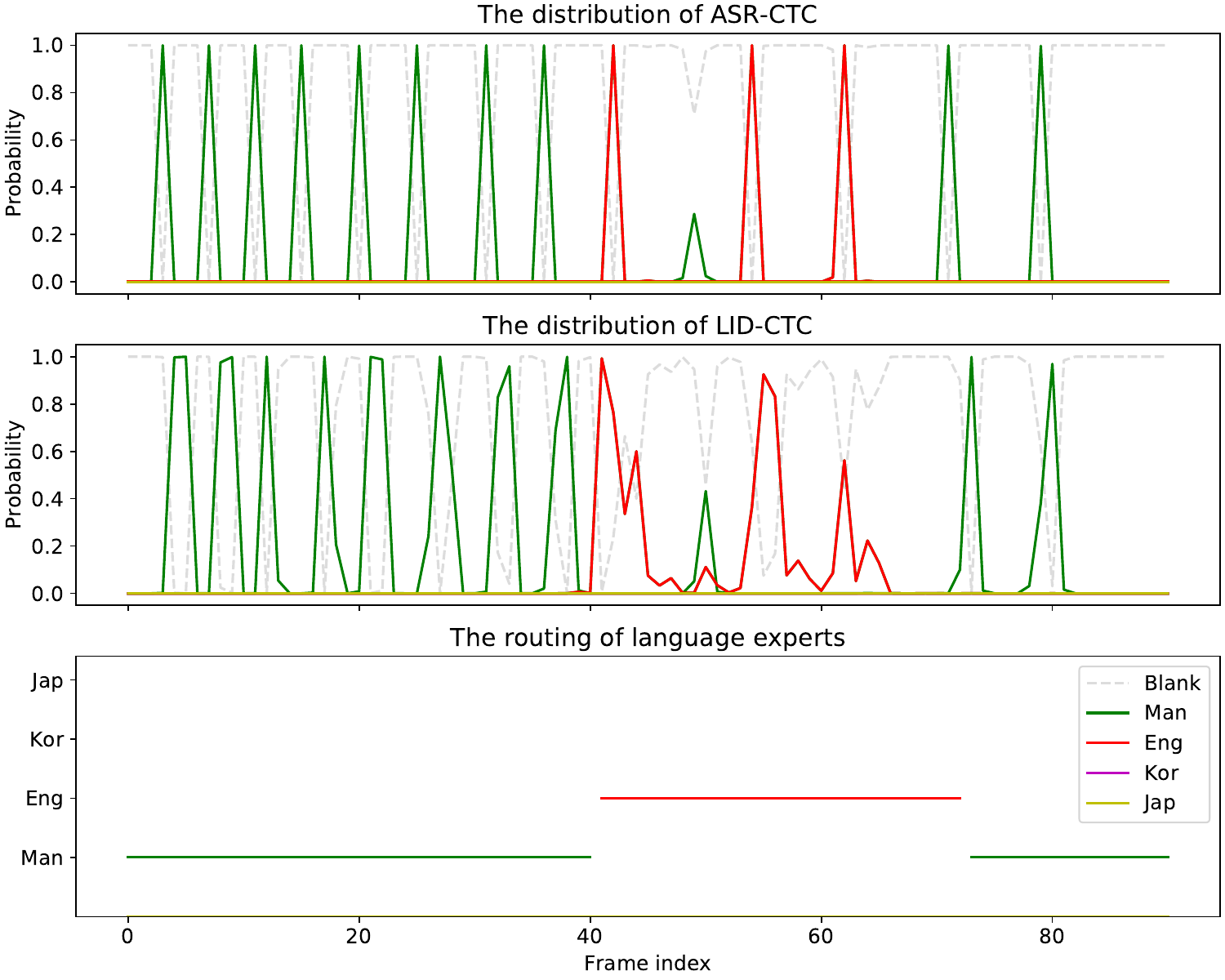}}
\caption{Visualization of the distribution of ASR-CTC, LID-CTC and language routing for the Mandarin-English code-switching speech.}
\vspace{-1.0 em}
\label{fig:cs-visual}
\end{figure}

As shown in Fig. \ref{fig:cs-visual}, for the code-switching input, the proposed method obtains the routing of the language expert by token-to-frame LID alignment and routes the layer inputs of the language segments to the corresponding language experts, which demonstrates the effectiveness of the proposed method for code-switching ASR.


\section{Conclusions}
\label{sec:conclusion}

This paper proposes the LR-MoE architecture to improve multilingual ASR system for monolingual and code-switching situations. Based on the frame-wise language-routing (FLR) mechanism, the proposed LR-MoE can switch the corresponding language expert block to extract language-specific representations adaptively and efficiently. Experiments show the LR-MoE significantly improves multilingual and code-switching ASR system over the standard Transformer model with comparable computational complexity and outperforms the previous MoE-based methods with less computational complexity. In the future, we will explore more efficient MoE routing mechanisms for multilingual and code-switching speech recognition.


\vfill
\pagebreak

\bibliographystyle{IEEEtran}
\bibliography{mybib}

\end{document}